\documentclass[conference,compsoc]{IEEEtran}
\ifCLASSOPTIONcompsoc
  \usepackage[nocompress]{cite}
\else
  \usepackage{cite}
\fi

\ifCLASSINFOpdf
\else
\fi

\usepackage{listings}
\usepackage{graphicx}
\usepackage{subcaption}
\usepackage{float}
\usepackage{rotating}
\usepackage{array}
\usepackage{enumitem}

\ifCLASSOPTIONcompsoc
  \usepackage[caption=false,font=footnotesize,labelfont=sf,textfont=sf]{subfig}
\else
  \usepackage[caption=false,font=footnotesize]{subfig}
\fi

\renewcommand{\paragraph}[1]{\vspace*{1pt plus 1pt minus .5pt}\noindent\textbf{#1}}

\begin{document}
%

\title{Aim High, Stay Private: Differentially Private Synthetic Data Enables Public Release of Behavioral Health Information with High Utility}




 \author{
   \IEEEauthorblockN{Mohsen Ghasemizade}
   \IEEEauthorblockA{Department of Computer Science\\
   University of Vermont\\
   mohsen.ghasemizade@uvm.edu}
   \and
   \IEEEauthorblockN{Juniper Lovato}
   \IEEEauthorblockA{Department of Computer Science\\
   University of Vermont\\
   juniper.lovato@uvm.edu}
     \and
   \IEEEauthorblockN{Christopher M. Danforth}
   \IEEEauthorblockA{Department of Mathematics and Statistics\\
   University of Vermont\\
   chris.danforth@uvm.edu}
   \and
   \IEEEauthorblockN{Peter Sheridan Dodds}
   \IEEEauthorblockA{Department of Computer Science\\
   University of Vermont\\
   peter.dodds@uvm.edu}
     \and
   \IEEEauthorblockN{Laura S. P. Bloomfield}
   \IEEEauthorblockA{The Gund Institute for the Environment\\
   University of Vermont\\
   laura.bloomfield@uvm.edu}
   \and
   \IEEEauthorblockN{Matthew Price}
   \IEEEauthorblockA{Department of Psychological Science\\
   University of Vermont\\
   matthew.price@uvm.edu}
   \and
   \IEEEauthorblockN{Team LEMURS}
   University of Vermont\\
   \and
   \IEEEauthorblockN{Joseph P. Near}
   \IEEEauthorblockA{Department of Computer Science\\
   University of Vermont\\
   jnear@uvm.edu}
 }


%


\maketitle
\thispagestyle{plain}
\pagestyle{plain}

\begin{abstract}
\boldmath
Sharing health and behavioral data raises significant privacy concerns, as conventional de-identification methods are susceptible to privacy attacks. Differential Privacy (DP) provides formal guarantees against re-identification risks, but practical implementation necessitates balancing privacy protection and the utility of data.

We demonstrate the use of DP to protect individuals in a real behavioral health study, while making the data publicly available and retaining high utility for downstream users of the data. We use the Adaptive Iterative Mechanism (AIM) to generate DP synthetic data for Phase 1 of the Lived Experiences Measured Using Rings Study (LEMURS). The LEMURS dataset comprises physiological measurements from wearable devices (Oura rings) and self-reported survey data from first-year college students. We evaluate the synthetic datasets across a range of privacy budgets, ${\epsilon} = {1}$ to ${100}$, focusing on the trade-off between privacy and utility.

We evaluate the utility of the synthetic data using a framework informed by actual uses of the LEMURS dataset. Our evaluation identifies the trade-off between privacy and utility across synthetic datasets generated with different privacy budgets. We find that synthetic data sets with ${\epsilon} = \mathbf{5}$ preserve adequate predictive utility while significantly mitigating privacy risks. 
Our methodology establishes a reproducible framework for evaluating the practical impacts of 
$\epsilon$ on generating private synthetic datasets with numerous attributes and records, contributing to informed decision-making in data sharing practices.
\unboldmath
\end{abstract}


%
\IEEEpeerreviewmaketitle

\section{Introduction}
The Lived Experiences Measured Using Rings Study (LEMURS) \cite{price2023large} 
recruited roughly 600 college students to explore the connections between well-being, health, heart activity, and sleep, employing Oura rings (a wearable device that measures heart rate and other physiological signals) and surveys. This study generated a dataset with more than 100 biometric measures and survey responses related to the participants' daily routines and demographics. While participants are assigned anonymous numerical IDs within the dataset, there is concern that certain column combinations could disclose the identities of some participants, particularly those in minority groups, such as transgender individuals and people of color.

Due to privacy concerns, datasets like these are typically not available to the public, and the LEMURS dataset is no exception. In recent years, a growing understanding of the weaknesses of traditional privacy solutions like de-identification~\cite{sweeney2024aboutmyinfo, ohm2009broken} has increased the barriers to sharing sensitive health-related data. For example, HIPAA's Safe Harbor provision prescribes a de-identification method~\cite{hhs2024hipaa}, but studies like LEMURS often do not consider these provisions strong enough to protect their participants.


Differential privacy (DP)~\cite{dwork2006calibrating, dwork2014algorithmic} is a formal mathematical framework for protecting the privacy of individuals that provides significantly better robustness against privacy attacks. DP mechanisms work by including randomization in the data generation process (often by adding random noise); this randomization provides strong privacy protection but can significantly reduce the utility of the data for downstream tasks. Navigating the \emph{privacy-utility tradeoff} remains a significant barrier for increased adoption of DP.

In this work, we demonstrate the feasibility of generating and releasing high-quality DP synthetic data for behavioral health studies like LEMURS. First, we show that even simple linkage attacks are sufficient to re-identify participants in the de-identified LEMURS data. Then, we generate and evaluate DP synthetic data for the LEMURS dataset that preserves the important properties of the original dataset, and we make the DP synthetic data available publicly.
To generate the synthetic data, we use the Adaptive Iterative Mechanism (AIM; \cite{mckenna2022aim}) with varying privacy budgets. We run AIM on a subset of the LEMURS data, to reduce dimensionality and improve utility, and we show that the synthetic data is robust against privacy attacks.

Evaluation of the utility of DP synthetic data is vital for navigating the privacy-utility tradeoff, but remains a major challenge---especially since the best utility metrics for a synthetic dataset depend on how the dataset is going to be used. We make progress towards resolving this challenge by performing a task-specific utility evaluation of our LEMURS synthetic data based on actual prior research findings on the non-DP data, and connecting the results to typical general-purpose utility metrics like workload error.


\paragraph{Contributions.}
In summary, we make the following contributions:

\begin{itemize}[leftmargin=14pt, itemsep=2pt]

\item We highlight the need for DP in behavioral health data by demonstrating successful privacy attacks against the de-identified LEMURS dataset.

  \item We demonstrate the feasibility of DP synthetic data in this setting by generating and releasing DP synthetic data for the LEMURS study. At a privacy budget of $\epsilon=5$, our synthetic data retains high utility.

\item We make progress toward guidance for navigating the privacy-utility tradeoff. We evaluate utility of our synthetic data against actual research uses of the original dataset, and draw connections between task-specific utility evaluation and more general metrics for utility.




\end{itemize}
The rest of the paper is organized as follows. Section~\ref{sec:backgr-relat-work} describes background and related work, including details of the LEMURS dataset. Section~\ref{sec:methodology} describes the methodology we used for conducting privacy attacks, generating DP synthetic data, and evaluating the utility of the synthetic data. Section~\ref{sec:results-analysis} presents our results, both on privacy (via attacks) and on utility. Section~\ref{sec:discussion} provides guidance for data releases based on our results and discusses open challenges.

\section{Background and Related Work}
\label{sec:backgr-relat-work}

\subsection{Differential Privacy}

Differential privacy (DP)~\cite{dwork2006calibrating, dwork2014algorithmic} is a widely adopted framework that safeguards the privacy of individuals within a dataset. By introducing controlled random noise drawn from a distribution, such as a Gaussian distribution, DP adds uncertainty to numerical columns. The magnitude of noise introduced into the data is governed by a privacy budget denoted as $\epsilon$. A DP mechanism ensures that the likelihood of producing any specific output from an input cannot fluctuate by more than a factor of $e^{\epsilon}$. Formally, a mechanism $\mathcal{M}$ satisfies DP if for any pair of neighboring datasets $D_1$ and $D_2$ and any possible set of outcomes $S$:
\begin{equation}
\label{eq:dp_definition}
\frac{\Pr[\mathcal{M}(D_1) \in S]}{\Pr[\mathcal{M}(D_2) \in S]} \leq e^{\epsilon}
\end{equation}
Intuitively, we can think of $\Pr[\mathcal{M}(D_1)]$ as the probability of outcome $S$ when an individual contributes their data, while $\Pr[\mathcal{M}(D_2)]$ denotes the probability of outcome $S$ when the individual does not contribute their data to the dataset. The DP definition bounds the ratio of these probabilities, ensuring that they must be similar---with $\epsilon$ controlling the degree of similarity. Smaller values of $\epsilon$ provide better privacy but generally result in less utility. Values of $\epsilon$ in the single digits are preferred for strong privacy.

A common relaxation of this definition is $(\epsilon, \delta)$-DP (also called approximate DP), which is defined similarly but has an additional parameter $\delta$:
\begin{equation}
\label{eq:dp_delta_definition}
\Pr[\mathcal{M}(D_1) \in S] \leq e^\epsilon \Pr[\mathcal{M}(D_2) \in S] + \delta
\end{equation}
The $(\epsilon, \delta)$-DP definition allows relaxing the guarantee for rare events, where $\delta$ controls how rare these events must be. The $\delta$ parameter is typically set very small, in a way that shrinks with the size of the dataset (e.g. $1/n^2$, where $n$ is the number of individuals in the data).

DP effectively prevents the re-identification of individuals within the dataset and has found extensive applications in medical settings to protect individual privacy \cite{dankar2012application, dwork2021promise, vu2009differential, zhang2012functional}. It is also beneficial in machine learning, as it helps safeguard attempts to infer whether a subject was included in the dataset used to train the model \cite{abadi2016deep}.

\subsection{Wearable Data}

Several recent studies have leveraged wearable devices such as the Oura Ring to monitor sleep, stress, and physiological patterns in various populations, including healthcare workers and college students \cite{fudolig2024collective}. While these studies have demonstrated the utility of wearable-derived data in predicting health outcomes, they also consistently raise concerns about privacy and the risk of re-identification. Researchers have emphasized the sensitive nature of continuous biometric data and the ethical responsibility to protect participant privacy, often implementing safeguards such as coded IDs, restricted data sharing, and secure storage. For example, large-scale efforts like UCSF's TemPredict study could not share raw data due to confidentiality agreements, and others noted that even de-identified datasets may still pose re-identification risks \cite{mason2022detection}. These concerns underscore the importance of developing more robust privacy-preserving techniques when working with wearable health data \cite{shiba2023assessing}.

\subsection{Privacy Attacks}

Privacy attacks exploit the fact that anonymized datasets can still be linked to external information, allowing individuals to be re-identified. These attacks have exposed serious vulnerabilities in traditional de-identification methods, especially when auxiliary data is readily available.

The Netflix Prize competition \cite{wikipedia2024netflixprize} aimed to improve movie recommendation algorithms by releasing anonymized user data. However, researchers were able to re-identify individual users by cross-referencing the Netflix dataset with IMDB ratings. In another instance, New York City's taxicab dataset, known as `Riding with the Stars' \cite{schmidt2014riding}, was re-identified by linking it with a celebrity gossip website data. This revealed private information about the travel habits of specific passengers. These incidents highlight how unexpectedly diverse auxiliary data sources, like IMDB or a celebrity gossip website, can be used for re-identification. Similarly, the U.S. Census Bureau has historically relied on methods like data suppression and noise addition to protect privacy, but these approaches often compromised data utility and failed to fully prevent re-identification \cite{dajani2017modernization}. Examples like these underscore the opportunity for DP to protect against unforeseen vulnerabilities.

Linkage attacks pose a significant and dangerous threat in medical settings, where each individual likely has data stored in multiple organizations and locations, and therefore will be included in various electronic medical record datasets. Furthermore, there may be overlapping rows and information from each individual in each of these settings. If one of these datasets is published, even after de-identification, an adversary could match the `anonymous' columns with an existing dataset or a dataset that only the adversary has access to, and re-identify individuals. Therefore, it is crucial to ensure that any information released, particularly for research purposes, is initially privately protected.

\subsection{DP Synthetic Data Generation}

One of the most appealing uses of DP is generating synthetic data --  records matching the input, intended to be broadly representative of the source data -- so that everyone can have access to the synthetic data for further analysis and studies, while the privacy of the individuals in the dataset is protected. 

Rosenblatt et al. \cite{rosenblatt2024epistemic} evaluated six prominent state-of-the-art methods for generating DP synthetic data, drawing inspiration from previous studies that demonstrated their effectiveness across diverse data release tasks \cite{mckenna2022aim, tao2021benchmarking}. These methods included MST \cite{mckenna2021winning},  AIM \cite{mckenna2022aim},  PrivMRF \cite{cai2021data},  PATECTGAN \cite{rosenblatt2020differentially},  PrivBayes \cite{zhang2017privbayes}, and  GEM \cite{liu2021iterative}.
MST, AIM, and PrivMRF are marginal-based methods that model low-dimensional relationships between variables. PrivBayes, on the other hand, is a Bayesian network-based method that remains competitive despite its age. PATECTGAN and GEM are deep learning-based synthesizers that provide substantial performance enhancements.

Each method possesses unique design strengths. Notably, AIM stands out for its flexibility and practical usability. PrivBayes constructs a Bayesian network and incorporates noise into k-way correlations, although it can be inefficient with high-dimensional data. MST constructs a private spanning tree to model relationships, but it lacks the ability to adapt to specific analysis objectives. AIM surpasses MST by being workload-aware and iteratively learning a synthetic distribution specifically tailored to a pre-selected set of key variables. This approach leads to significantly improved utility. PrivMRF attempts to optimize marginal selection, yet it still faces complexity challenges as datasets expand. Deep learning-based methods such as PATECTGAN incur substantial model training costs and exhibit reduced efficiency for smaller datasets \cite{rosenblatt2024epistemic}.

AIM uses a graphical model to construct a maximum spanning tree among attributes in the data feature space, where edges are weighted by mutual information \cite{mckenna2021hdmm}. AIM relies on this algorithm for parameterizing the underlying distribution \cite{mckenna2022aim}, and it is workload aware, meaning it parametrizes a private synthetic distribution through an iterative process, taking advantage of a set of queries of interest, or priorities. AIM is one of the recent state-of-the-art DP synthetic data generators and well-maintained implementations are available \cite{rosenblatt2024epistemic, nist_privacy_archive}.

\subsection{Evaluation of the Utility of DP Results}

Evaluating utility in DP results is challenging. Tabular releases are often evaluated using simple accuracy metrics like L1 and L2 error, or relative error variants. Synthetic data is typically evaluated by measuring error between a set of 2-way or 3-way marginals calculated on the original and synthetic datasets. These approaches are independent from any particular downstream use of the data, and may not reflect the utility of the data for an actual real-world use.

Recently, Rosenblatt et al. \cite{rosenblatt2024epistemic} introduced the concept of epistemic parity as a novel lens for assessing the efficacy of DP synthetic datasets. Instead of solely focusing on statistical metrics such as marginal L1 and L2 errors, or correlation preservation, they contend that synthetic data's utility should be gauged by whether the scientific conclusions derived from the original data remain reproducible after applying DP. In their evaluation, they specifically reconstruct pivotal analyses from published papers, often regression models, and compare critical attributes, such as the significance of regression coefficients, between the real and private datasets. Their proposed epistemic parity metric evaluates not only raw statistical similarity but also whether substantive findings and relationships are faithfully preserved.

Although our study primarily employs conventional metrics, measuring marginal L1 and L2 errors, computing Spearman correlation heat maps, and inspecting structure through PCA \cite{jolliffe2002principal} and UMAP \cite{mcinnes2018umap}, the underlying principle of epistemic parity motivates the deeper objective of our regression model evaluation.
Our methodology evaluates whether synthetic datasets retain the same intervariable associations observed in real data. This is achieved through correlation analyses across key behavioral and survey features. While our methods adhere to standard evaluation metrics, they also implicitly assess whether fundamental scientific relationships in the health study are preserved under DP.

\subsection{LEMURS Dataset} 

The Lived Experiences Measured Using Rings Study (LEMURS) \cite{lemurs_project} was conducted from the fall of 2022 to the spring of 2025 at the University of Vermont. Over 600 first-year college students participated in this longitudinal experiment to assess changes in students' sleep, stress, mental health, and other outcomes through a series of weekly surveys. Weekly assessments included self-report surveys such as the Perceived Stress Scale and the Generalized Anxiety Disorder 7-Item Scale (GAD-7), as well as demographics, stressful events, and other questions. The survey dataset includes a variety of categorical variables, such as responses to Likert-scale and multiple-choice questions. To enable numerical analysis, all categorical responses were encoded as numeric values. For instance, a question with possible answers \textit{``Low,'' ``Medium,''} and \textit{``High''} is represented in the dataset as 0, 1, and 2, respectively. A separate codebook defines the meaning of each encoded value, ensuring traceability throughout the analysis.

Additionally, researchers utilized Oura rings, smart rings that track various health and wellness metrics, to collect Total Sleep Time (TST), resting heart rate, heart rate variability, sleep stages (REM, deep, light), and other physiological data from the participants.

The primary goal of LEMURS is to monitor and improve the mental health and well-being of college students during a critical life transition, with a particular focus on identifying links between objective physiological signals and mental health outcomes. Researchers aimed to predict stress levels or mental health deterioration before symptoms become severe, using sleep patterns and wearable-derived metrics to detect early signs of risk. Prior work using the LEMURS dataset has demonstrated these connections. For example, Bloomfield et al. \cite{bloomfield2024predicting} found associations between sleep metrics and perceived stress using mixed-effects models, while another study \cite{bloomfield2023events} identified behavioral predictors of anxiety symptoms such as social media use and poor sleep quality. Fudolig et al. \cite{fudolig2024two} showed that delayed heart rate recovery during sleep was linked to trauma and anxiety symptoms. Together, these findings highlight the value of combining passive physiological monitoring with self-reported surveys to enable scalable mental health interventions. Moreover, releasing differentially private versions of these datasets can support inclusive research on vulnerable populations, such as minority or marginalized groups, without compromising individual privacy.

Our work uses two datasets collected in the LEMURS study. The first dataset, drawn from Phase 1 of the LEMURS project, contains 108 columns with a mix of qualitative and quantitative features. The second dataset is more compact, consisting of 19 purely quantitative columns that include physiological measurements from the Oura ring along with self-reported Perceived Stress Scale scores. For simplicity, we refer to the larger dataset as \textbf{the survey dataset} and the smaller one, composed almost entirely of Oura-based measurements, as \textbf{the Oura dataset}.

\section{Methodology}
\label{sec:methodology}

To assess the generated datasets with varying values of epsilon, we first introduce two attack scenarios to demonstrate that the AIM-generated synthetic datasets are resistant to these attacks. Subsequently, we introduce diverse evaluation methods to illustrate the trade-off between utility and privacy. Sections~\ref{sec:linkage-attack} and~\ref{sec:memb-infer-attack} describe the privacy attacks we use to evaluate privacy, and Section~\ref{sec:util-eval-meth} describes our approaches for evaluating utility of the DP synthetic data.

\subsection{Linkage Attack}
\label{sec:linkage-attack}

To evaluate the privacy of the original LEMURS dataset, we conducted a linkage attack. A linkage attack is a privacy attack in which an adversary uses external data sources, such as public records or survey data, to re-identify individuals in an anonymized dataset. Even when direct identifiers like names or email addresses are removed, quasi-identifiers (such as age, sleep duration, or stress scores) can still be used to match individuals across datasets and uncover sensitive information.

In the study by Bloomfield et al. \cite{bloomfield2024predicting}, the authors used Oura Ring data and survey responses from the LEMURS dataset to build a mixed-effects linear regression model. Their goal was to examine how total sleep time and the week of the semester affected perceived stress scores. 

Assuming the Oura dataset from Bloomfield et al. as the target dataset, and the survey dataset as an auxiliary dataset available to an adversary, a linkage attack becomes feasible. Since both datasets contain overlapping attributes, such as total sleep time and perceived stress scores, an attacker could compare those features and successfully re-identify individuals in the Oura dataset. However, DP prevents this type of attack by introducing carefully calibrated noise into the data or query outputs. DP offers a strong and quantifiable guarantee against linkage attacks by maintaining uncertainty about individual contributions, even in the presence of auxiliary information \cite{balle2020hypothesis, near2023guidelines, wood2018differential}. 

\subsection{Membership Inference Attack via Closest Distance}
\label{sec:memb-infer-attack}

To evaluate the privacy of our DP synthetic data, we conducted a membership inference attack. Our goal is to test whether an attacker, seeing only the differentially private synthetic data, can tell if a particular individual was included in the original training set.  We use a Closest Distance approach from TAPAS, a Toolbox for Adversarial Privacy Auditing of Synthetic Data \cite{houssiau2022tapas}. Specifically, we extended the TAPAS library by integrating the AIM synthetic data generator, allowing us, for the first time, to comprehensively evaluate AIM-generated synthetic datasets under TAPAS's attack framework.

The attacker knows the target's characteristic values and holds a 50\% random sample of the real data for calibration. They repeatedly ask the AIM mechanism for synthetic draws of the same size, and in each draw, they find the single synthetic record that is closest to the target (using either Hamming or Euclidean distance). Intuitively, if the target was part of the training data, its synthetic twin will tend to appear closer in the release. By collecting these `nearest neighbours' distances from draws where the target was included versus excluded, the attacker builds an ROC curve and selects the distance threshold that best separates `in' from `out'. Finally, with that threshold fixed, a fresh synthetic draw is used to decide membership: if the nearest neighbor is within the threshold, the attacker guesses `in' otherwise `out'.

\subsection{Utility Evaluation Methods}
\label{sec:util-eval-meth}

Based on the survey for Phase 1 of the study, the objective was to create a synthetic, private version of the dataset that would allow researchers to access it for further studies while preserving individual privacy and maintaining its utility. To preserve utility, we selected the AIM algorithm to generate the synthesized data. Using various privacy budgets ($\epsilon = [1, 2, 5, 10, 20, 50, 100]$), we generated six different versions of the survey dataset and six versions of the Oura dataset. 

The algorithm employed in AIM is constrained to generating synthetic data for quantitative values due to the inherent nature of DP, which introduces numerical noise into numerical data. Consequently, all synthesized data generated through DP comprises solely numerical columns. Qualitative columns, such as text-based responses regarding reasons for removing the ring, enjoyment derived from using the ring, and locations of activities, have been excluded from the survey dataset.

\subsubsection{Regression models}

Building on the analysis by Bloomfield et al. \cite{bloomfield2024predicting}, which used a mixed-effects regression model to predict perceived stress scores from features such as total sleep time, we first replicated their model on the original data and confirmed that we could recover their published coefficients. We then applied the same model to each synthetic dataset produced by the AIM algorithm, using replication fidelity as a case-study metric.

In addition to this replication, we developed a second model using random forest regression to predict stress levels in the larger survey dataset. To select input features, we calculated Spearman correlation scores with perceived stress scores and selected the top 12 most correlated variables. These included reported Generalized Anxiety Disorder scores, caffeine consumption patterns, sleep duration, time spent outdoors, and relaxation time and other related columns. We chose a random forest model because of its ability to capture non-linear relationships and its robustness to noise, both of which are particularly beneficial when working with this dataset. To assess model performance, we used the $R^2$ score, which ranges from 0 (no predictive power beyond the mean) to 1 (perfect prediction). Higher $R^2$ values indicate stronger preservation of predictive utility in the synthetic datasets.

\subsubsection{Spearman Correlation}

To establish a baseline understanding of relationships between features in the dataset, we use Spearman correlation, a rank-based metric that is well-suited for the types of data in this study. Unlike Pearson correlation, which assumes linear relationships and interval-scaled variables, Spearman correlation captures monotonic trends, making it more appropriate for ordinal data such as survey responses or ranked scores. Additionally, Spearman is robust to outliers, offering more stable estimates of association in noisy or non-normally distributed data, conditions commonly encountered in behavioral and physiological datasets. This makes it a valuable tool for assessing whether key correlations are preserved across original and differentially private synthetic datasets.

\subsubsection{UMAP}

Uniform Manifold Approximation and Projection (UMAP) \cite{mcinnes2018umap} is a nonlinear dimensionality reduction technique that enables visualization of high-dimensional data in two or three dimensions. In this study, where we work with a high-dimensional survey dataset (108 columns) and the Oura dataset (19 columns), UMAP is an effective tool for comparing the structural integrity of the original data with its differentially private synthetic counterparts. By projecting these datasets into a 2D space, we can visually assess the extent of structural distortion introduced by differential privacy. UMAP also uncovers latent cluster structures, enabling us to examine whether individuals with similar attributes, such as comparable perceived stress scores, remain grouped together after the application of DP. This unsupervised clustering approach is particularly useful for evaluating latent patterns, similar to how Ghasemizade et. al \cite{ghasemizade2024developing} leveraged hierarchical models to uncover groupings in complex, belief-driven datasets.

\subsubsection{Marginal distribution metrics: L1 and L2 distances}

To quantify how well the synthetic datasets preserve the original data's structure, we compute marginal distribution errors using both L1 and L2 distance metrics. These metrics compare the frequency distributions of attribute combinations, specifically 2-way marginals between the real and synthetic datasets. L1 distance (also known as Manhattan distance) provides a straightforward measure of total absolute difference between distributions, while L2 distance (Euclidean distance) emphasizes larger discrepancies due to its squaring of differences. By averaging these errors across all marginal queries, we can assess how closely the synthetic data approximates the real data's joint distributions, which is especially important for preserving meaningful relationships in downstream analysis. These metrics are commonly used in the literature as task-agnostic measures of utility for differentially private data release.

\section{Results and Analysis}
\label{sec:results-analysis}


We first present the results of our empirical privacy analysis via attacks (Sections~\ref{sec:linkage-attack-1} and~\ref{sec:memb-infer-attack-1}). Then, we present results from our utility analysis. In Section~\ref{sec:regression-models}, we attempt to replicate regression results from prior research using the LEMURS dataset on our DP synthetic data. In Section~\ref{sec:spearman-correlation}, we use Spearman correlation heatmaps to show that correlations between attributes are generally maintained in our DP synthetic data. Section~\ref{sec:umap} evaluates the quality of the DP synthetic data by clustering with UMAP, and Section~\ref{sec:l1-l2-distances} evaluates data quality using the traditional approach of measuring L1 and L2 error on marginal workloads.

\subsection{Linkage Attack}
\label{sec:linkage-attack-1}

We assessed the similarity between the original Oura dataset and the survey dataset. With the two datasets in hand, and knowing the matching columns `week' and `perceived stress scores (PSS)', we looked for which rows exactly match in that respect. To make the attack robust, we added a third condition, namely the similarity score of reported sleep time from the survey dataset and the measured sleep time with the Oura ring. Since they are not expected to be the same, we allow some error between the two values in each dataset. 

Using the `Record Linkage Toolkit' Python package \cite{record_linkage}, we compared the numeric `total sleep time (TST)' values between the original Oura and survey datasets using a linear similarity function. We configure this function with an offset of 0.5, meaning that if the difference in sleep hours between the two datasets is within $\pm$0.5 hours, it is considered a perfect match (similarity score = 1.0).

Beyond this threshold, the similarity score decreases linearly with a scale parameter of 1.5, which controls the rate of decay. This configuration allows for partial credit when sleep values differ by more than 0.5 hours but still remain close. For instance, a difference of 1.0 hour results in a similarity score of approximately 0.67, and a difference of 2.0 hours yields a score near zero.

\begin{lstlisting}[
    caption={Configuration of Linkage Attack Using \texttt{recordlinkage}},
    label={lst:linkage-config},
    basicstyle=\ttfamily\footnotesize,
    breaklines=true,
    frame=single,
    columns=fullflexible
]
# Comparison configuration
compare.numeric('TST', method='linear', offset=0.5, origin=7, scale=1.5)

# Match condition: exact match on week and PSS, similarity > 0.7 for sleep
matches = compare_vectors[
    (compare_vectors['Week'] == 1) &
    (compare_vectors['PSS'] == 1) &
    (compare_vectors['TSS'] > 0.8)]
\end{lstlisting}

Using the configuration shown above, we identified four matching rows between the original Oura and survey datasets based on exact matches for week and perceived stress scores, along with a high similarity threshold for reported sleep duration. These matches demonstrate a successful linkage attack under relatively simple assumptions. We will use these four identified records in the next subsection for the membership inference attack on differentially private versions of the datasets. However, if the auxiliary dataset had been released under DP, the added noise would have prevented such confident re-identification, as DP ensures that no individual record can be reliably linked, even when some attributes overlap across datasets.

\subsection{Membership Inference Attack via Closest Distance}
\label{sec:memb-infer-attack-1}

To evaluate the risk of membership inference on the differentially private synthetic data, we applied the Closest Distance attack to eight total records identified via linkage, four records each from the original Oura and original Survey datasets. Due to computational constraints, specifically the expense of repeatedly generating synthetic data for each target, we performed this evaluation only at a privacy budget of $\epsilon = 10$, using a single AIM synthesizer trained on each full dataset. The targeted records are labeled ID1–ID4 for the Oura dataset and ID5–ID8 for the survey dataset. The fact that $\epsilon = 10$ still defeats the attack suggests that $\epsilon =5$ is actually a fairly conservative choice for the privacy budget in this case.

\subsubsection{Oura Dataset}

\begin{figure}[t]
  \centering
  \includegraphics[width=\linewidth]{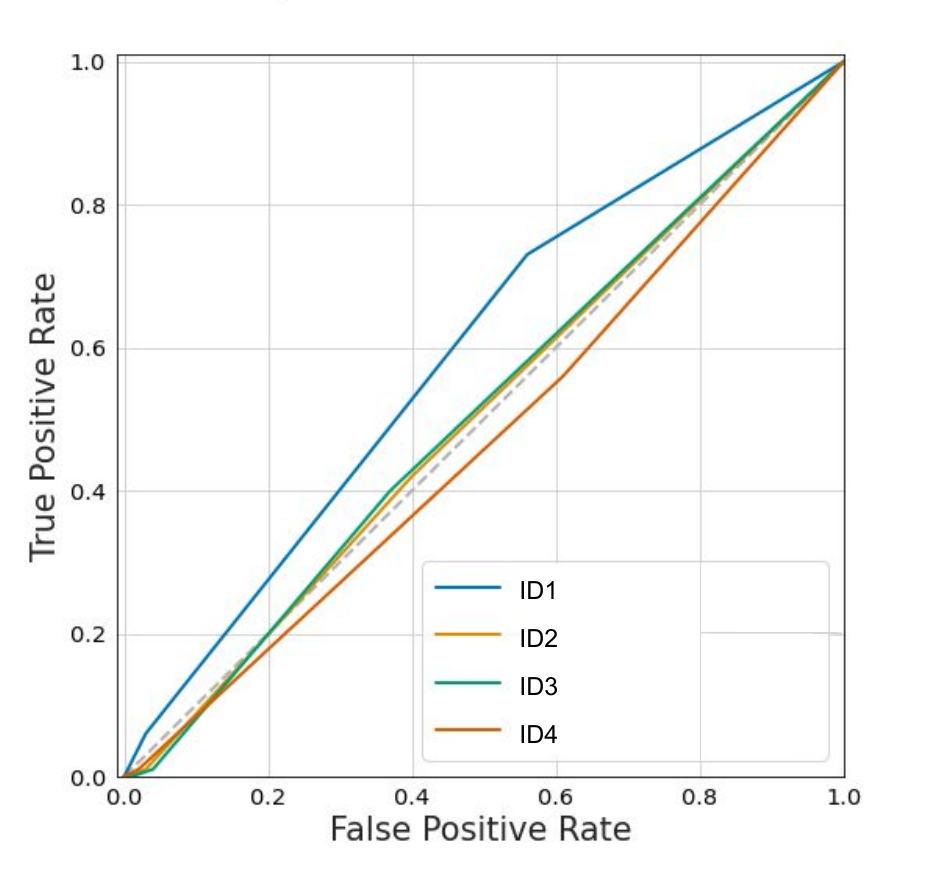}
  \caption{ROC curves for the Closest‐Distance membership inference attack on four linkable Oura records (ID1–ID4) at \(\epsilon = 10\).}
  \label{fig:oura-mia-roc}
\end{figure}

The ROC curves resulting from the Closest Distance Membership Inference attack on the four targets on DP Oura dataset are shown in \ref{fig:oura-mia-roc}. Target 2, 3, and 4 produced ROC curves nearly indistinguishable from random guessing, and Target 1 showed a marginally better true- positive rate. This indicates that the DP Oura dataset with $\epsilon = 10$, although it contains easily identifiable records through linkage, effectively prevents membership inference attacks at $\epsilon = 10$. These results reinforce the practicality and strength of the preservation of privacy of this chosen privacy budget.

\subsubsection{Survey Dataset}

\begin{figure}[t]
  \centering
  \includegraphics[width=\linewidth]{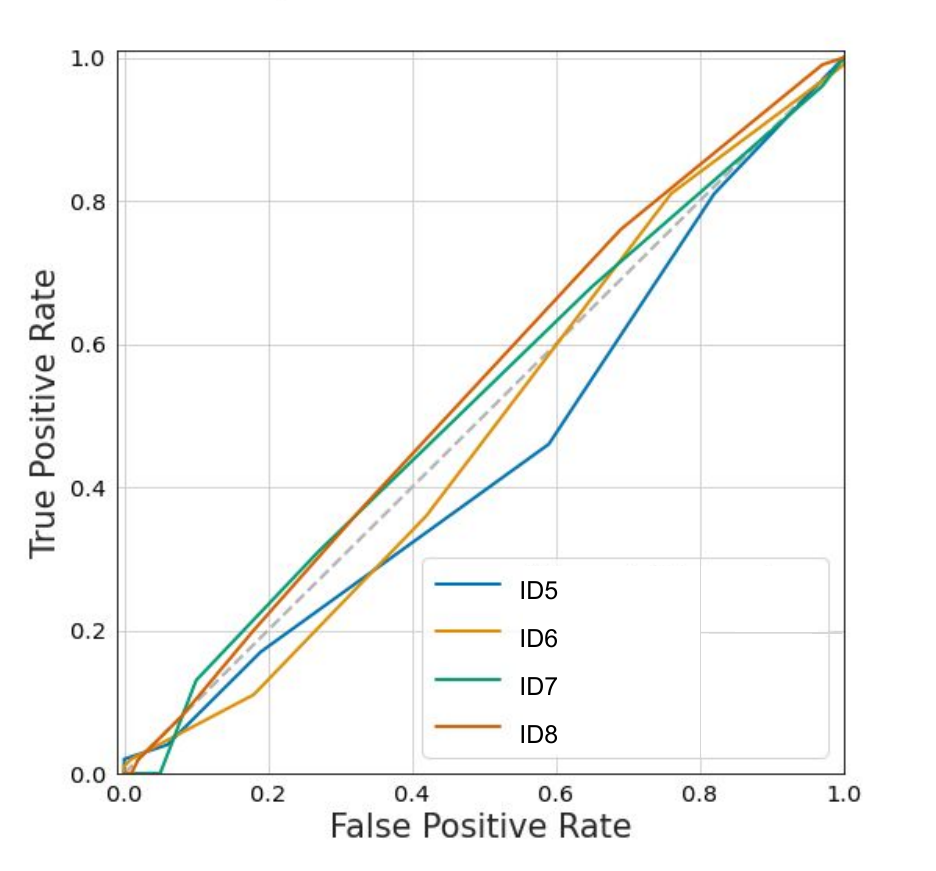}
  \caption{ROC curves for the Closest‐Distance membership inference attack on four linkable survey records (ID5–ID8) at \(\epsilon = 10\).}
  \label{fig:survey-mia-roc}
\end{figure}

Figure~\ref{fig:survey-mia-roc} presents the ROC curves for the Closest-Distance attack on the four Survey dataset targets. Targets 6, 7, and 8 closely follow the diagonal random baseline with only minimal deviations. Target 5, however, is below the random baseline, indicating that the attack was less effective than random guessing. These slight deviations do not represent meaningful privacy risks, confirming that the Survey dataset is robust against membership inference at $\epsilon = 10$. Thus, our results consistently support the conclusion that $\epsilon= 10$ offers effective protection against practical privacy threats in this dataset.

\subsection{Regression Models}
\label{sec:regression-models}

As previously discussed, we developed two regression models for the two distinct datasets we have: the Oura dataset and the survey dataset. For each dataset, we created an original model without noise and six distinct models for each of the DP-generated synthetic datasets designed for each epsilon.

\subsubsection{Oura Dataset}

To replicate the mixed-effects regression model used in Bloomfield et al. \cite{bloomfield2024predicting}, we first converted the Oura Ring total sleep time from seconds to hours for consistency with the survey dataset. We also created a centered version of total sleep time by subtracting each participant's mean sleep duration (tst\_dev) to account for within-person variation. We then fit a linear mixed-effects model predicting perceived stress scores as a function of week, gender, and within-participant total sleep time deviation, with participant ID as a random effect and week included in the random slope. This model serves as a reference to evaluate how well the synthetic datasets preserve these key associations. The coefficients for `week' and `tst\_dev' are outlined in Table~\ref{tab:epsilon-coefficients}. In the original model, the coefficient for \texttt{week} is $-0.331$, indicating that perceived stress scores tend to decrease slightly as the semester progresses, while the coefficient for \texttt{tst\_dev} is $-0.897$, suggesting that students who sleep less than their personal average report significantly higher stress levels. Table 1 also shows the computed coefficients for different synthetic datasets with varying epsilons.

\begin{table}
\centering
\caption{Comparison of \texttt{week} and \texttt{tst\_dev} Coefficients Across $\epsilon$ Values}
\label{tab:epsilon-coefficients}
\begin{tabular}{ccc}
\hline
\textbf{$\epsilon$} & \textbf{week} & \textbf{tst\_dev} \\
\hline
Original & -0.331 & -0.897 \\
1   & -0.048 & 0.130 \\
2   & -0.156 & -0.028 \\
\textbf{5}   & \textbf{-0.210} & \textbf{-0.316} \\
\textbf{10}  & \textbf{-0.334} & -0.214 \\
\textbf{20}  & -0.270 & \textbf{-0.944} \\
50  & -0.404 & -0.576 \\
100 & -0.460 & -0.679 \\
\hline
\end{tabular}
\end{table}

When comparing the regression coefficients across different levels of DP, we see that 
$\epsilon = 10$ still yields the week coefficient (–0.210) closest to the original (–0.331), and $\epsilon = 20$ matches the \texttt{tst\_dev} coefficient (–0.944) most closely. However, the $\epsilon = 5$ synthetic dataset produces coefficients of –0.210 for week and –0.316 for \texttt{tst\_dev}, representing a reasonable compromise between utility and privacy. Although $\epsilon = 5$ does not align as precisely with either original parameter as $\epsilon = 10$ or $\epsilon = 20$, it still preserves the overall negative trend (stress decreases over time and increases as sleep falls below an individual’s mean). This suggests that $\epsilon = 5$ maintains useful predictive structure while offering stronger privacy protection. In practice, choosing $\epsilon = 5$ can be preferable when an analyst is willing to accept a modest utility loss in order to substantially reduce disclosure risk.

\subsubsection{Survey Dataset}

Using the random forest model described in Section 5.2.1, we evaluated the predictive performance across both real and synthetic versions of the survey dataset. As shown in Table~\ref{tab:r2-rf}, the $R^2$ score of the random forest model trained on the original survey dataset is 0.710. Among the synthetic datasets, the best-performing result was observed at $\epsilon = 50$ with an $R^2$ of 0.726. However, this high utility comes at the cost of reduced privacy. Notably, the synthetic dataset generated with $\epsilon = 5$ achieved an $R^2$ score of 0.680, which is not far from the original performance while maintaining a significantly lower privacy budget. This result highlights the classic privacy-utility tradeoff: while higher $\epsilon$ values yield better utility, lower values like $\epsilon = 5$ still offer strong predictive performance with better privacy guarantees due to the addition of more noise.

\begin{table}
\centering
\caption{R$^2$ Scores for Random Forest Regression Across $\epsilon$ Values}
\label{tab:r2-rf}
\begin{tabular}{cc}
\hline
\textbf{$\epsilon$} & \textbf{$R^2$ Score} \\
\hline
Original & 0.710 \\
1   & 0.386 \\
2   & 0.606 \\
\textbf{5}   & \textbf{0.680} \\
10  & 0.693 \\
20  & 0.641 \\
\textbf{50}  & \textbf{0.726} \\
100 & 0.733 \\
\hline
\end{tabular}
\end{table}

\subsection{Spearman Correlation}
\label{sec:spearman-correlation}

To visualize the pairwise relationships between variables, we generated Spearman correlation heatmaps for both the original and synthetic versions of the Oura and survey datasets. These heatmaps allow us to assess whether key correlation structures are preserved after applying DP.

\subsubsection{Oura Dataset}

\begin{figure*}
\centering
\includegraphics[height=.9\textheight]{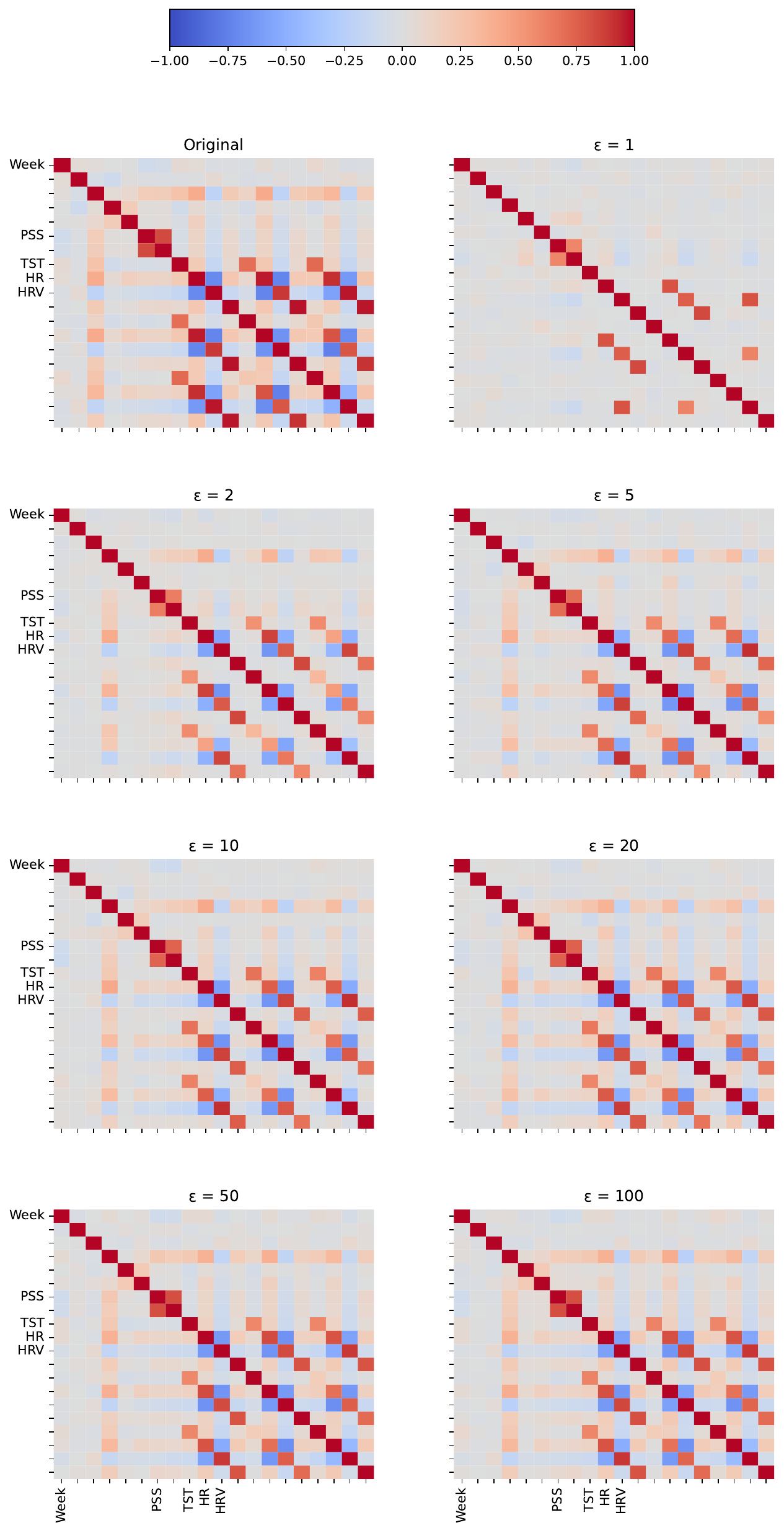}
\caption{Spearman correlation heatmaps for the original and synthetic Oura datasets across various $\epsilon$ levels. Only selected variable labels are shown to improve readability. HR = Heart Rate, PSS: Perceived Stress Score, TST: Total Sleep Time, HRV: Heart Rate Variability.}
\label{fig:oura-custom-labels}
\end{figure*}

Figure~\ref{fig:oura-custom-labels} presents Spearman correlation heatmaps for the original Oura dataset and its differentially private synthetic versions across various $\epsilon$ values. As expected, the heatmap for $\epsilon = 1$, which corresponds to the highest noise level, shows a significantly different structure compared to the original, with weaker and less consistent correlation patterns. In contrast, the heatmaps for higher $\epsilon$ values more closely resemble the original. While some variation in the strength of individual correlations is visible (with certain cells appearing slightly stronger or weaker), the overall structure and grouping of correlated variables are largely preserved. This suggests that the synthetic datasets maintain meaningful relationships between key physiological features at modest to high privacy budgets.

\subsubsection{Survey Dataset}

\begin{figure*}
\centering
\includegraphics[height=.9\textheight]{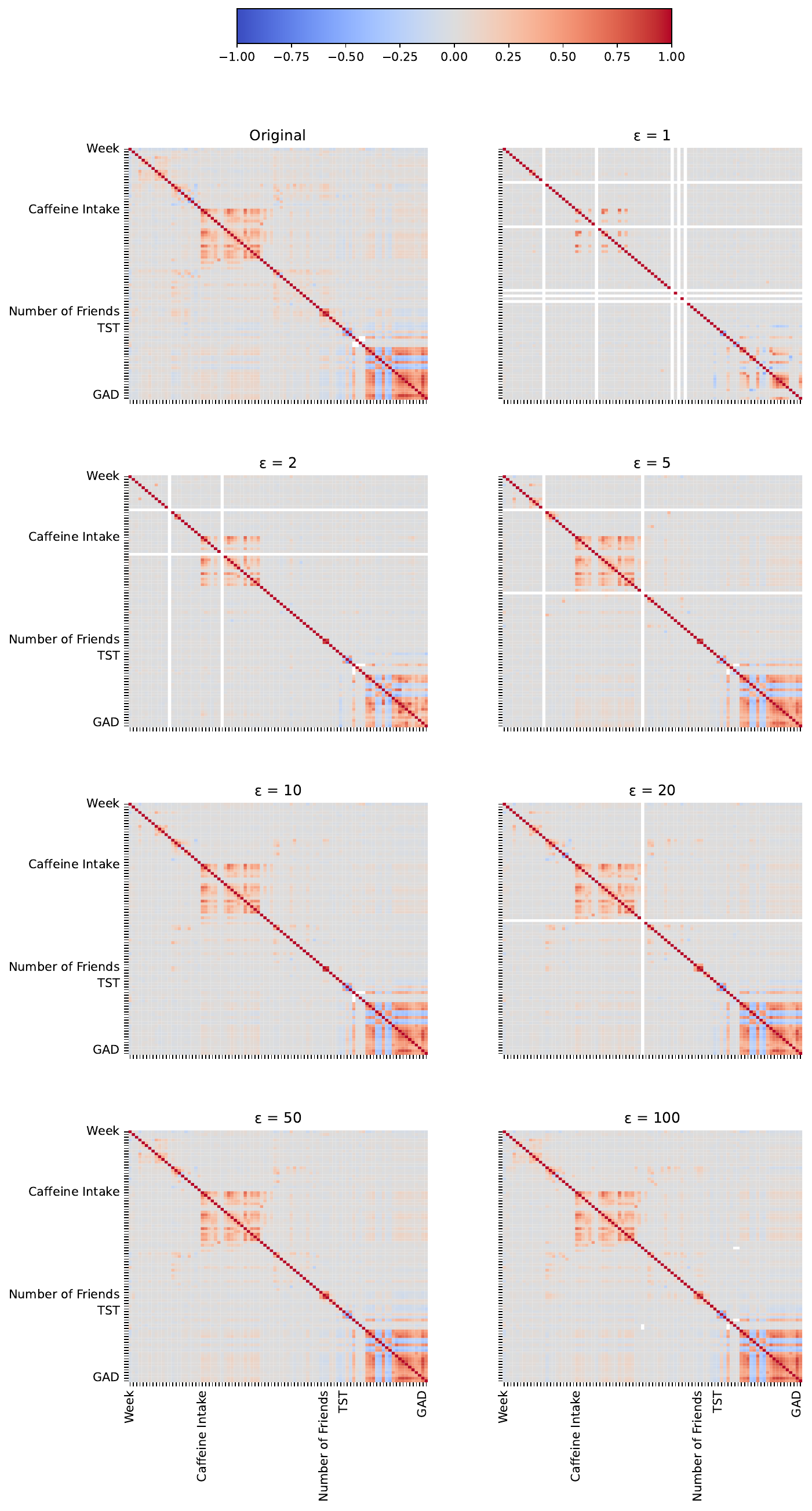}
\caption{Spearman correlation heatmaps for the original and synthetic survey datasets at various $\epsilon$ values. All variables are included, but only selected axes labels are shown for clarity: Week, PSS: Perceived Stress Score, GAD: Generalized Anxiety Disorder, TST: Total Sleep Time, caffeine intake, and number of friends.}
\label{fig:survey-heatmaps}
\end{figure*}

Figure~\ref{fig:survey-heatmaps} shows Spearman correlation heatmaps for the original survey dataset and its synthetic counterparts generated with varying $\epsilon$ values. As expected, the heatmap for $\epsilon = 1$ demonstrates significant distortion due to high noise, leading to weak or broken correlation structure across most variables. In contrast, datasets generated with higher privacy budgets ($\epsilon \geq 5$) better preserve the original correlation patterns. Notably, the bottom-right cluster of variables, which includes PSS: Perceived Stress Score, GAD: Generalized Anxiety Disorder, and stress-related features, shows consistent structure across $\epsilon = 5$, $10$, $20$, $50$, and $100$. Among these, the heatmaps for $\epsilon = 5$ and $\epsilon = 10$ most closely resemble the original, suggesting that these privacy levels strike an effective balance between privacy protection and statistical utility for this dataset.

\subsection{UMAP}
\label{sec:umap}

The behavior of UMAP is governed by a few key hyperparameters, most notably `n\_neighbors' and `min\_dist'. The `n\_neighbors' parameter balances local versus global structure by controlling the number of neighboring points used to estimate the manifold structure. Lower values emphasize local detail and can fragment clusters, while higher values promote more globally consistent embeddings. The `min\_dist' parameter controls how tightly UMAP packs points together in the low-dimensional space. Smaller values lead to denser, more compact clusters, whereas larger values encourage more spread and separation between groups. These hyperparameters were tuned empirically to produce interpretable visualizations that reflect both the continuity of perceived stress score and the topological preservation of structure across synthetic datasets. UMAP is particularly effective on high-dimensional data, such as our 108 column survey dataset.

\subsubsection{Oura Dataset}

\begin{figure}
\centering
\setlength{\tabcolsep}{1pt}

\begin{tabular}{ccc}
\begin{minipage}{0.21\textwidth}
  \begin{tabular}{@{}c@{}}
    \begin{tabular}{cc}
      \rotatebox{90}{\small (a) Original} &
      \includegraphics[width=0.9\linewidth]{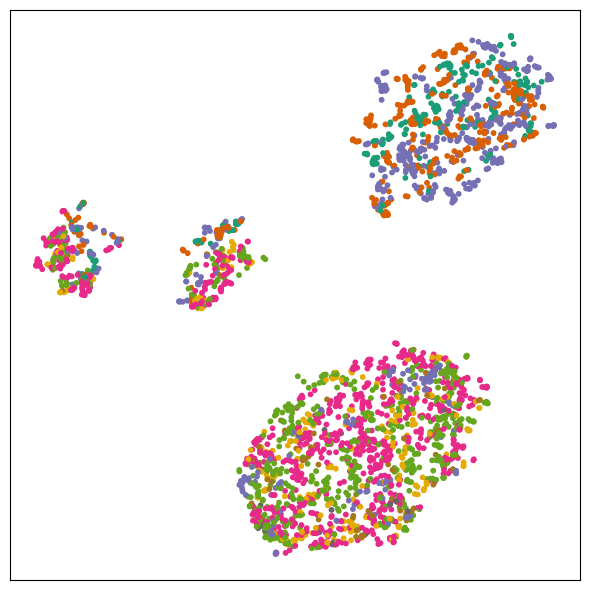}
    \end{tabular} \\[1ex]
    \begin{tabular}{cc}
      \rotatebox{90}{\small (c) $\epsilon = 2$} &
      \includegraphics[width=0.9\linewidth]{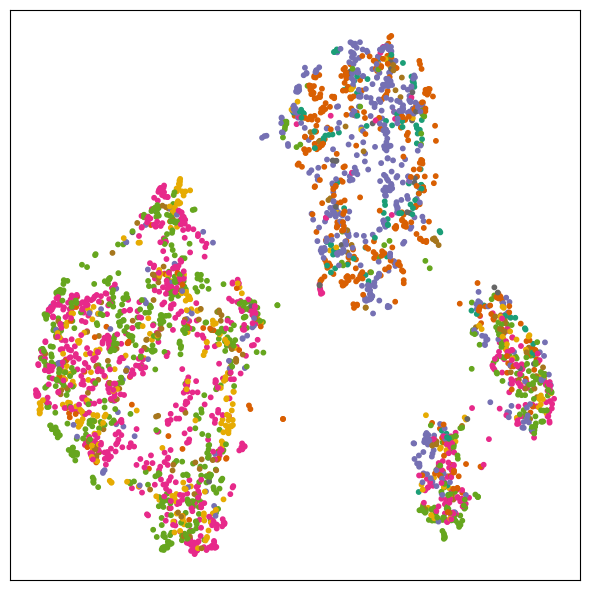}
    \end{tabular} \\[1ex]
    \begin{tabular}{cc}
      \rotatebox{90}{\small (e) $\epsilon = 10$} &
      \includegraphics[width=0.9\linewidth]{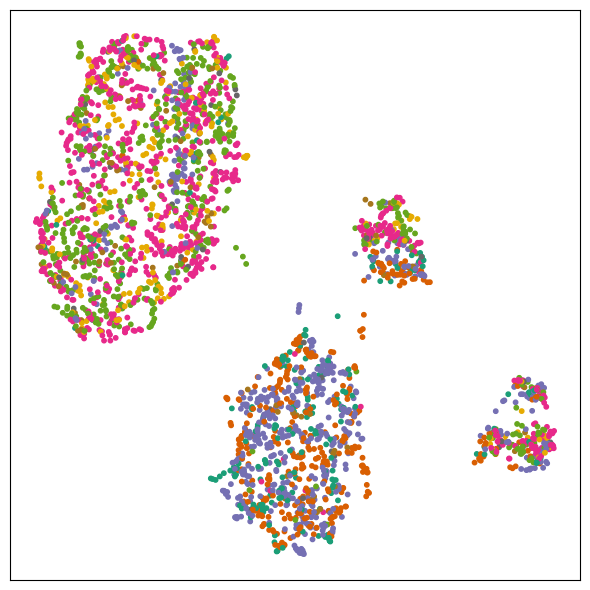}
    \end{tabular} \\[1ex]
    \begin{tabular}{cc}
      \rotatebox{90}{\small (g) $\epsilon = 50$} &
      \includegraphics[width=0.9\linewidth]{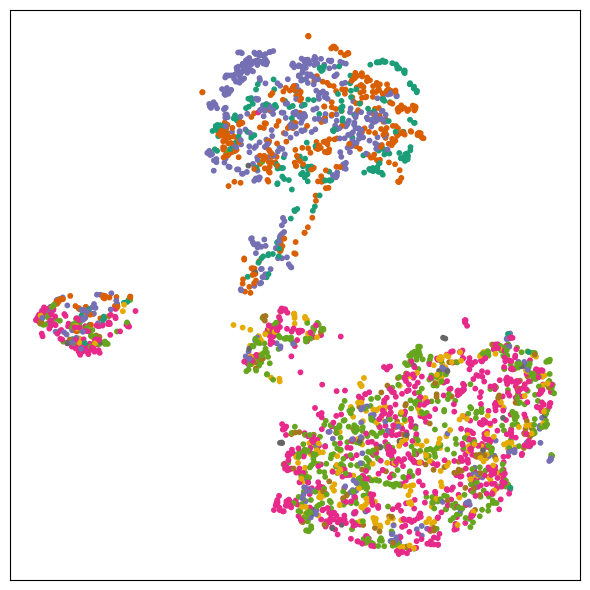}
    \end{tabular}
  \end{tabular}
\end{minipage}
&
\begin{minipage}{0.21\textwidth}
  \begin{tabular}{@{}c@{}}
    \begin{tabular}{cc}
      \includegraphics[width=0.9\linewidth]{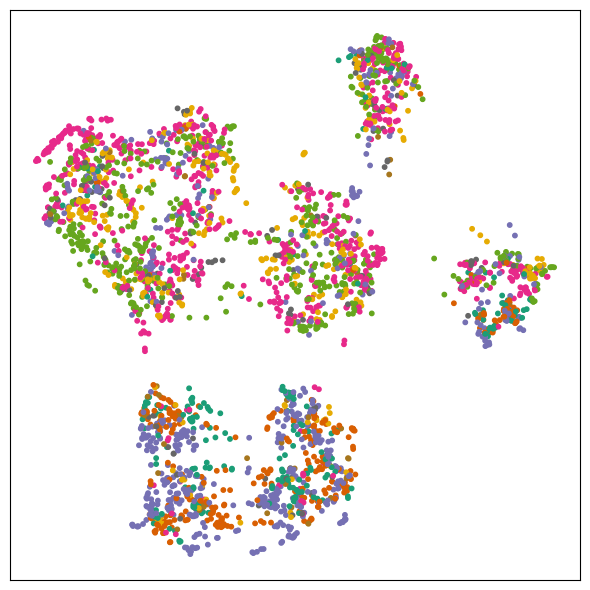} &
      \rotatebox{90}{\small (b) $\epsilon = 1$}
    \end{tabular} \\[1ex]
    \begin{tabular}{cc}
      \includegraphics[width=0.9\linewidth]{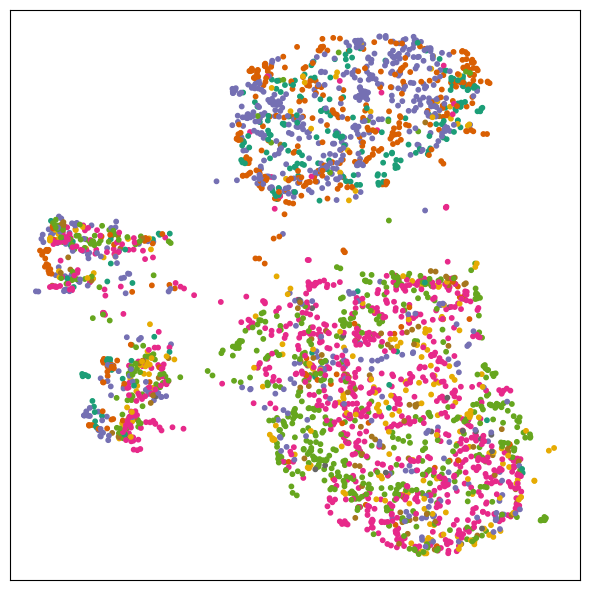} &
      \rotatebox{90}{\small (d) $\epsilon = 5$}
    \end{tabular} \\[1ex]
    \begin{tabular}{cc}
      \includegraphics[width=0.9\linewidth]{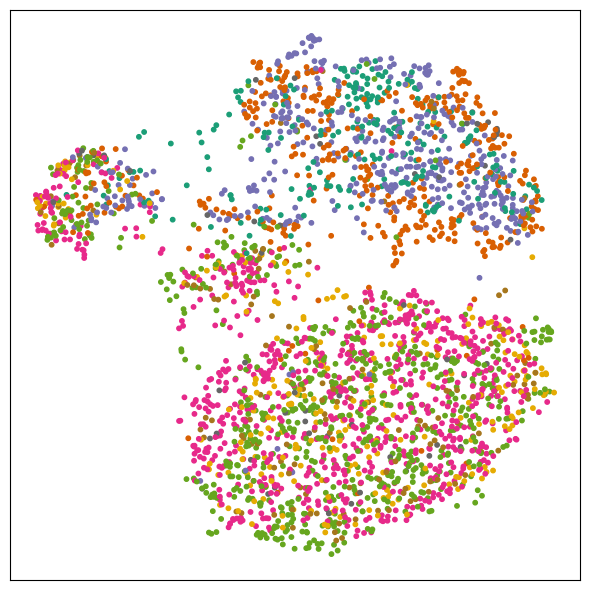} &
      \rotatebox{90}{\small (f) $\epsilon = 20$}
    \end{tabular} \\[1ex]
    \begin{tabular}{cc}
      \includegraphics[width=0.9\linewidth]{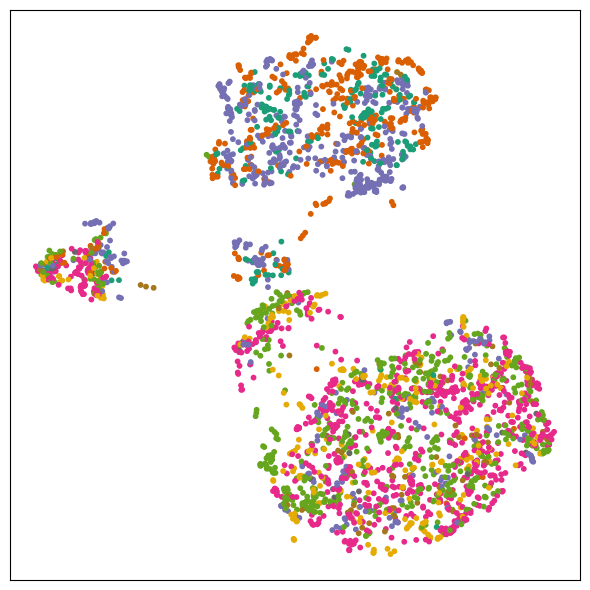} &
      \rotatebox{90}{\small (h) $\epsilon = 100$}
    \end{tabular}
  \end{tabular}
\end{minipage}
&
\begin{minipage}{0.10\textwidth}
  \includegraphics[width=\linewidth]{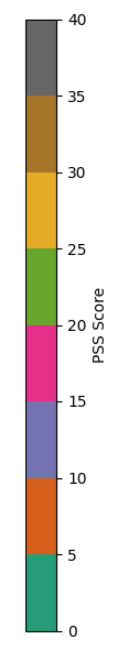}
\end{minipage}
\end{tabular}

\caption{UMAP projections of the original and synthetic Oura datasets for different $\epsilon$ values. Vertical captions are placed beside each image, and the legend on the right shows perceived stress scores (PSS).}
\label{fig:oura-umap}
\end{figure}

To observe the structural fidelity of the synthetic Oura datasets, we projected the original and DP versions into 2D using UMAP with \texttt{n\_neighbors=45} and \texttt{min\_dist=0.025}, shown in Figure~\ref{fig:oura-umap}. This configuration prioritizes global coherence while allowing for tight, well-defined local clusters. The choice of a very small \texttt{min\_dist} (0.025) is particularly suitable here because the Oura dataset is relatively low-dimensional (19 columns), and the data shows naturally compact clusters in its latent structure. 

In the original dataset, we observe a clear separation into four primary clusters. The top left cluster predominantly consists of people with low perceived stress score, ranging from 0 to 15. The remaining three clusters are populated mostly with higher perceived stress score, particularly in the 15–20 range, with some individuals reaching scores up to 40. Among the synthetic datasets, the version generated with $\epsilon= 5$ and $10$, show the strongest structural resemblance to the original. The cluster geometry and distribution of perceived stress level values are highly consistent with the original projection, suggesting that $\epsilon= 5$, $10$ strike a compelling balance between utility and privacy in the Oura dataset.

\subsubsection{Survey Dataset}

To evaluate the structure of the synthetic survey datasets, we applied UMAP with \texttt{n\_neighbors = 65} and \texttt{min\_dist = 0.6}. These hyperparameters were chosen to emphasize more global relationships and reduce over-fragmentation in a high-dimensional dataset (108 columns), which tends to benefit from stronger global preservation. A higher \texttt{min\_dist} encourages UMAP to maintain greater distance between points in the lower-dimensional embedding, making the overall clustering structure more interpretable.

The original dataset reveals a well-defined structure with approximately five clusters predominantly representing perceived stress score in the 15--20 range (pink), five clusters around the 20--25 range (green), and a distinct yellow cluster corresponding to higher stress levels (30--35). A smaller grouping near the center includes mostly perceived stress score between 10--15 (purple). Among the synthetic datasets, the projections for $\epsilon = 5$, $\epsilon = 10$ and $\epsilon = 100$ most closely resemble the original structure, preserving both the spatial arrangement and perceived stress scores color distribution of the clusters. In contrast, lower $\epsilon$ values (e.g., $\epsilon = 1$ or $\epsilon = 2$) lead to noticeable dispersion and less consistent cluster geometry, indicating a loss of utility from strong privacy constraints.

\begin{figure}
\centering
\setlength{\tabcolsep}{1pt}

\begin{tabular}{ccc}
\begin{minipage}{0.21\textwidth}
  \begin{tabular}{@{}c@{}}
    \begin{tabular}{cc}
      \rotatebox{90}{\small (a) Original} &
      \includegraphics[width=0.9\linewidth]{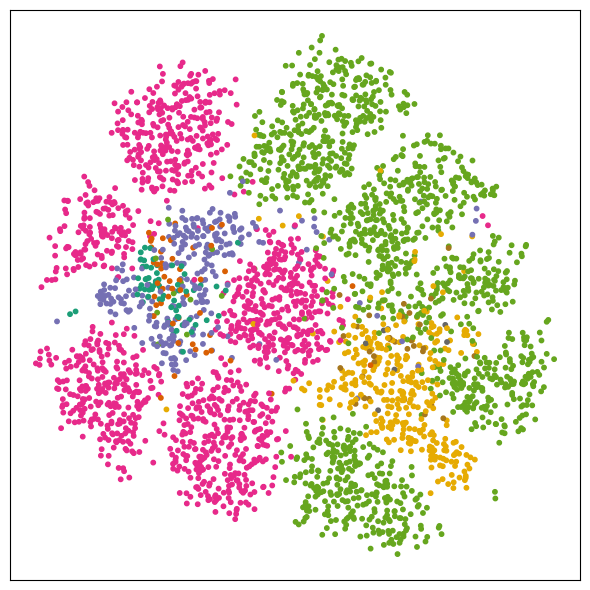}
    \end{tabular} \\[1ex]
    \begin{tabular}{cc}
      \rotatebox{90}{\small (c) $\epsilon = 2$} &
      \includegraphics[width=0.9\linewidth]{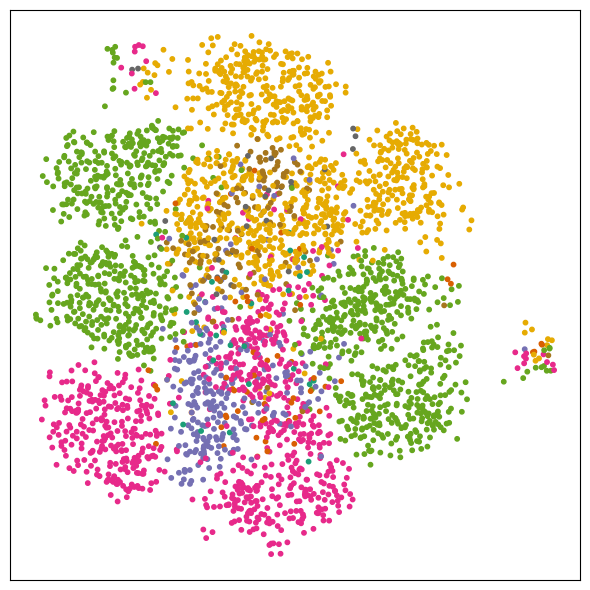}
    \end{tabular} \\[1ex]
    \begin{tabular}{cc}
      \rotatebox{90}{\small (e) $\epsilon = 10$} &
      \includegraphics[width=0.9\linewidth]{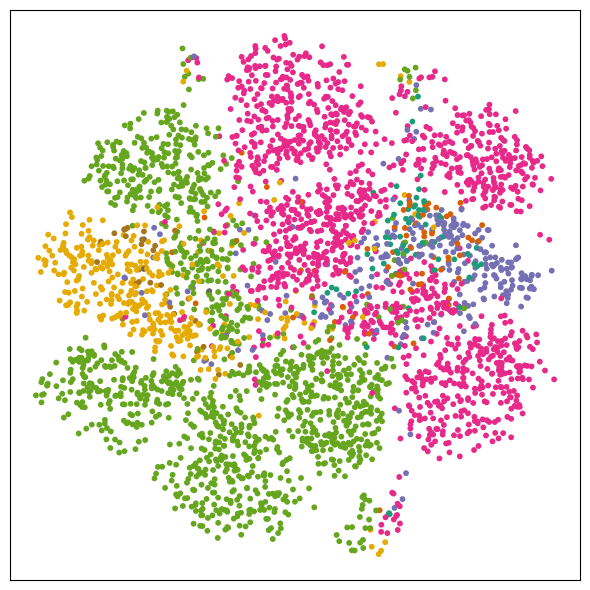}
    \end{tabular} \\[1ex]
    \begin{tabular}{cc}
      \rotatebox{90}{\small (g) $\epsilon = 50$} &
      \includegraphics[width=0.9\linewidth]{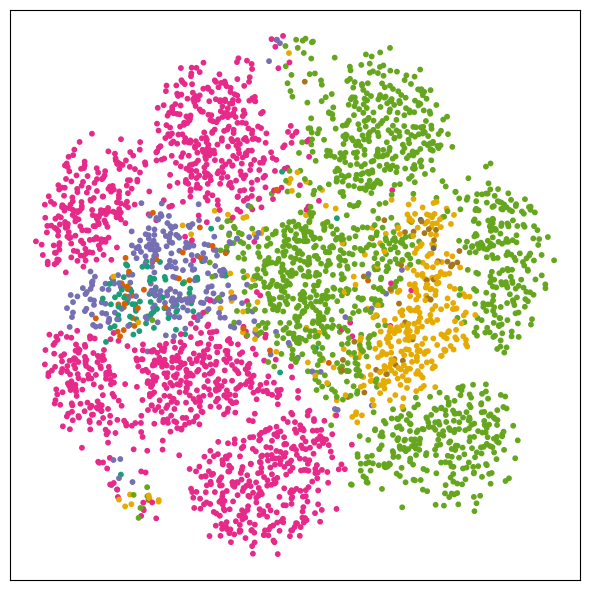}
    \end{tabular}
  \end{tabular}
\end{minipage}
&
\begin{minipage}{0.21\textwidth}
  \begin{tabular}{@{}c@{}}
    \begin{tabular}{cc}
      \includegraphics[width=0.9\linewidth]{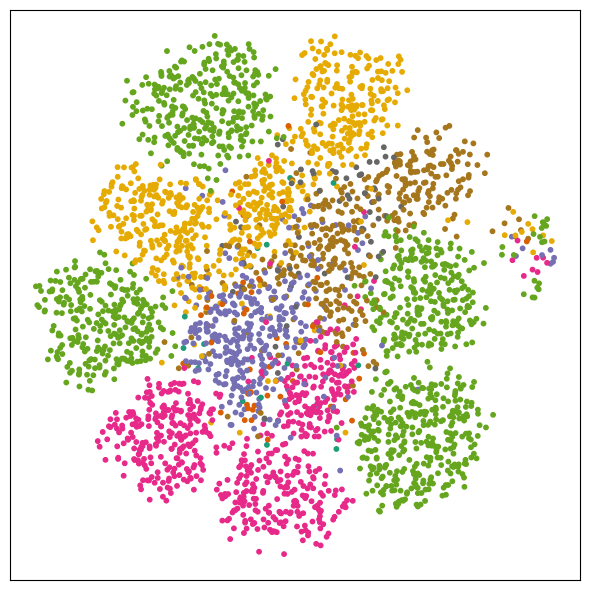} &
      \rotatebox{90}{\small (b) $\epsilon = 1$}
    \end{tabular} \\[1ex]
    \begin{tabular}{cc}
      \includegraphics[width=0.9\linewidth]{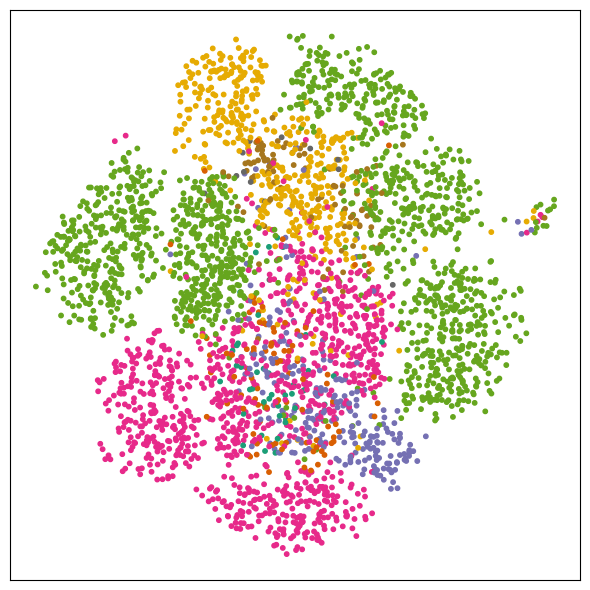} &
      \rotatebox{90}{\small (d) $\epsilon = 5$}
    \end{tabular} \\[1ex]
    \begin{tabular}{cc}
      \includegraphics[width=0.9\linewidth]{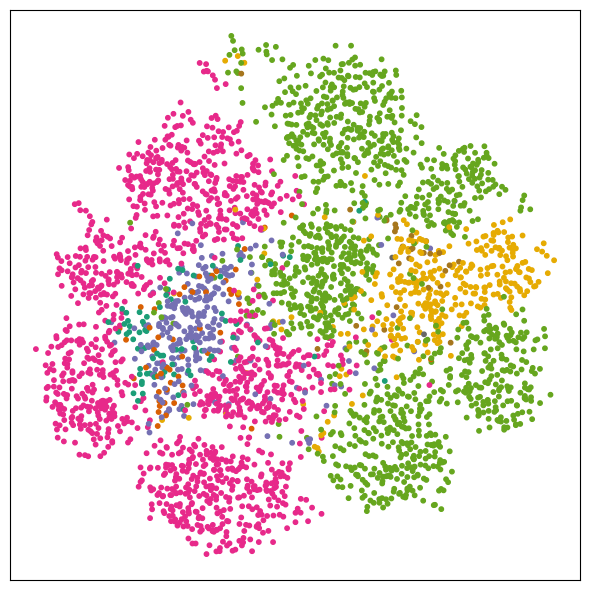} &
      \rotatebox{90}{\small (f) $\epsilon = 20$}
    \end{tabular} \\[1ex]
    \begin{tabular}{cc}
      \includegraphics[width=0.9\linewidth]{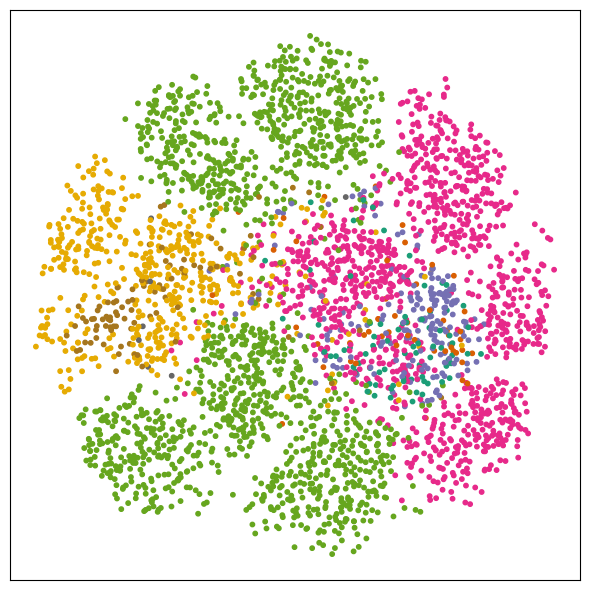} &
      \rotatebox{90}{\small (h) $\epsilon = 100$}
    \end{tabular}
  \end{tabular}
\end{minipage}
&
\begin{minipage}{0.10\textwidth}
  \includegraphics[width=\linewidth]{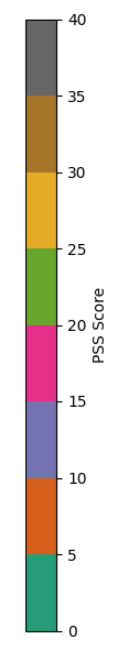}
\end{minipage}
\end{tabular}

\caption{UMAP projections of the original and synthetic survey datasets for different $\epsilon$ values. Vertical captions are placed beside each image, and the legend on the right shows perceived stress score (PSS).}
\label{fig:survey-umap}
\end{figure}

\subsection{L1 and L2 Error on Marginal Workloads}
\label{sec:l1-l2-distances}

Inspired by the approach in \cite{mckenna2022aim}, which evaluates synthetic data quality by computing the L1 and L2 distances of 2- or 3-way marginal queries between the real and synthetic datasets, we adopt a similar evaluation strategy. Specifically, we measure the L1 and L2 distances for all 2-way marginals between the original and synthetic versions of both the Oura and survey datasets as an indicator of utility preservation.

\subsubsection{Oura Dataset}

\begin{figure}
\centering
\begin{subfigure}{0.48\textwidth}
    \centering
    \includegraphics[width=\linewidth]{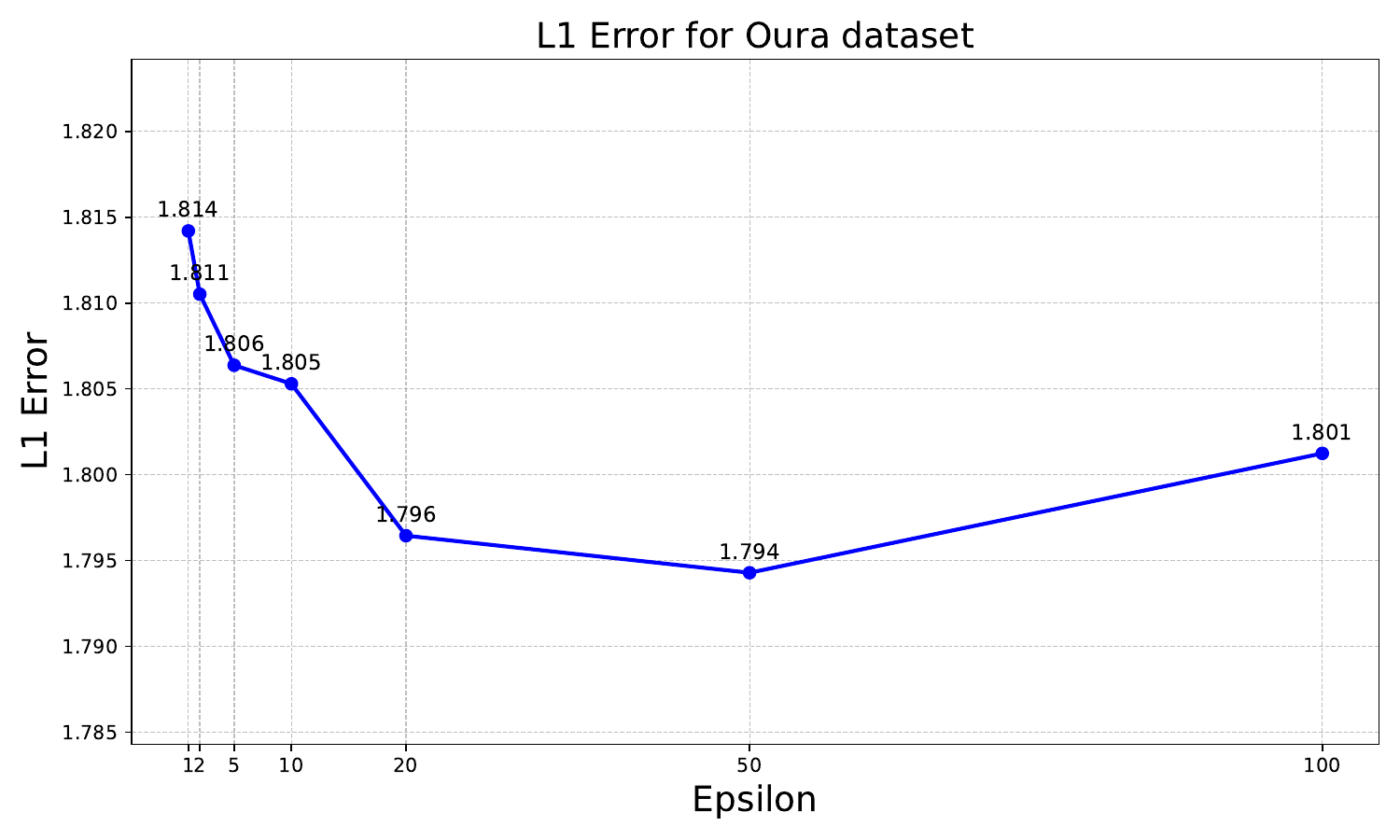}
    \label{fig:l1-survey}
\end{subfigure}
\hfill
\begin{subfigure}{0.48\textwidth}
    \centering
    \includegraphics[width=\linewidth]{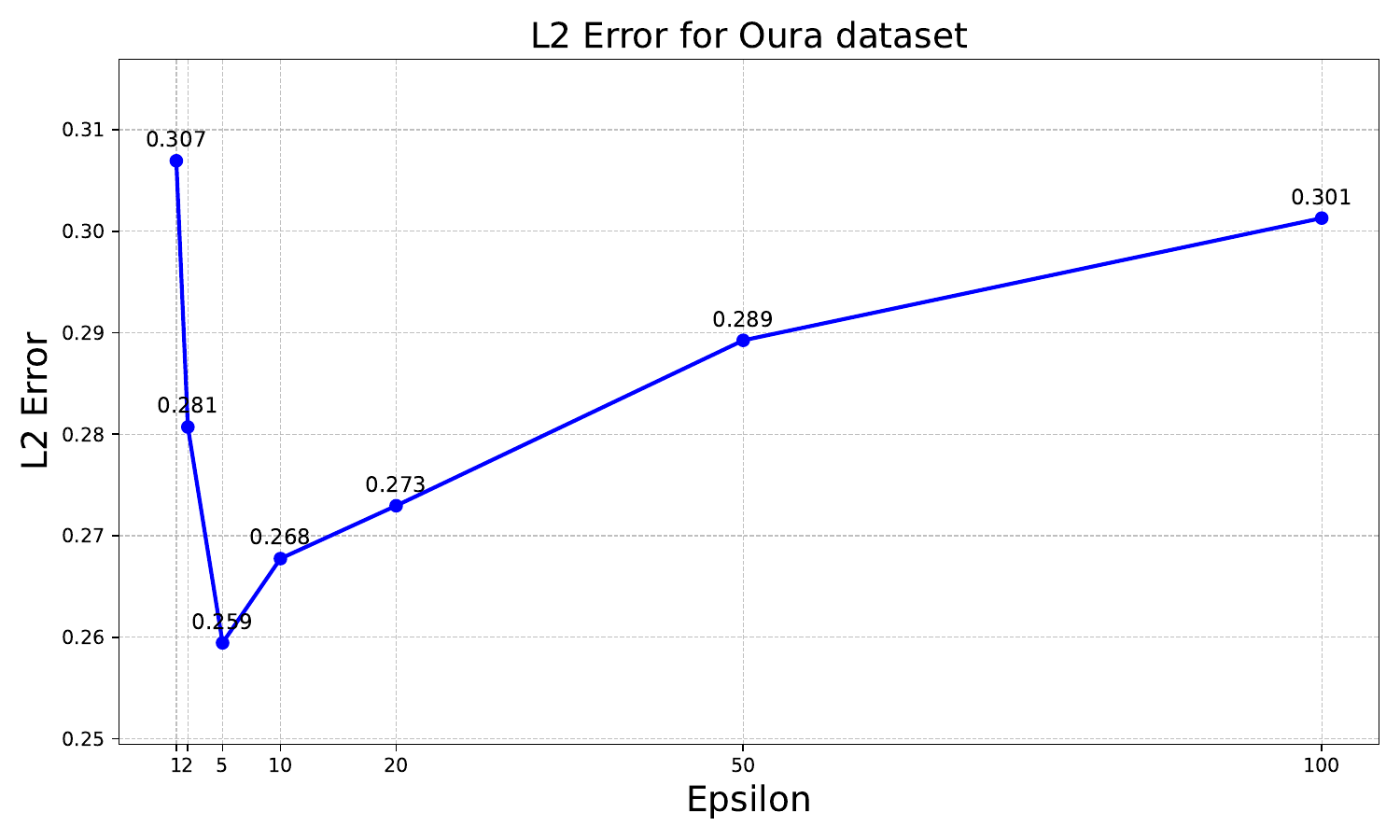}
    \label{fig:l1-oura}
\end{subfigure}
\caption{L1  and L2 marginal errors across different $\epsilon$ values for the Oura dataset. The y-axis is in logarithmic scale to reflect small variations in error.}
\label{fig:oura-errors}
\end{figure}

In the Oura dataset, our findings, shown in Figure~\ref{fig:oura-errors}, indicate that the synthetic data generated with an $\epsilon$ value of 50 achieves the lowest L1 error of 1.794, which is relatively close to the L1 error obtained when $\epsilon$ is set to 5, resulting in an error of 1.806. The L2 error is minimized for $\epsilon$ equal to 5, with a value of 0.259.

\subsubsection{Survey Dataset}

\begin{figure}
\centering
\begin{subfigure}{0.48\textwidth}
    \centering
    \includegraphics[width=\linewidth]{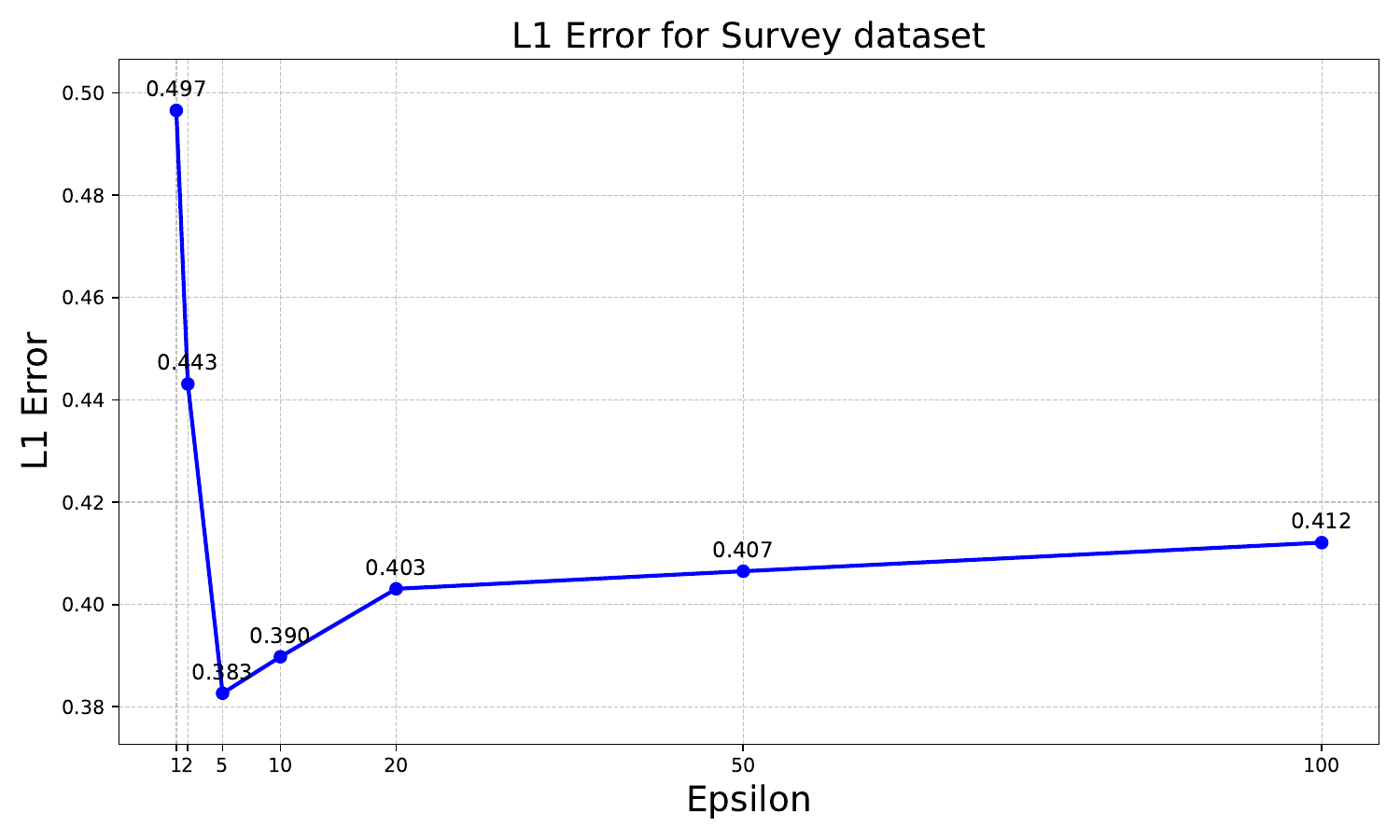}
    \label{fig:l2-survey}
\end{subfigure}
\hfill
\begin{subfigure}{0.48\textwidth}
    \centering
    \includegraphics[width=\linewidth]{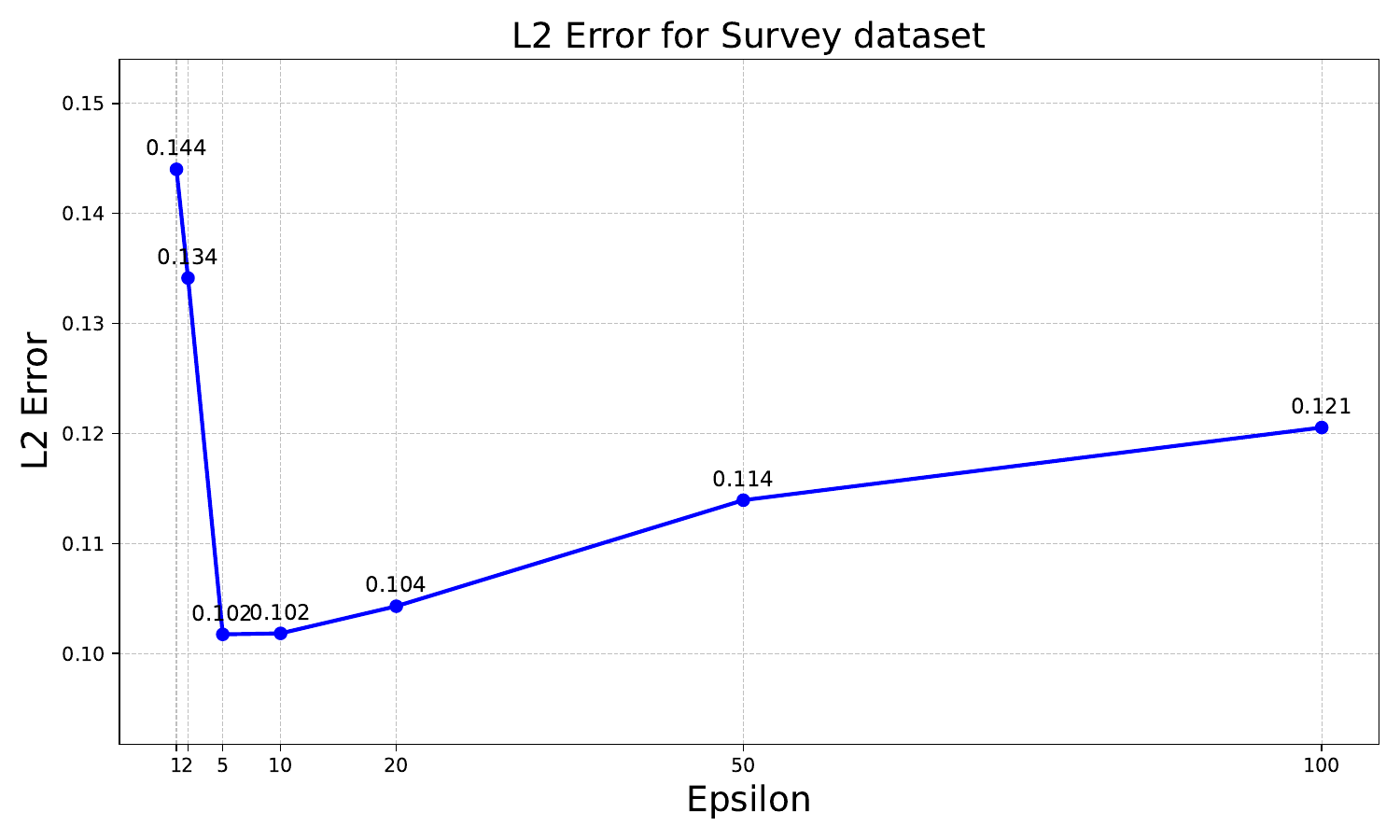}
    \label{fig:l2-oura}
\end{subfigure}
\caption{L1  and L2 marginal errors across different $\epsilon$ values for the survey dataset. The y-axis is on a logarithmic scale to reflect small variations in error.}
\label{fig:survey-errors}
\end{figure}

In the survey dataset, the L1 error reaches its minimum, shown in Figure~\ref{fig:survey-errors} at $\epsilon = 5$. The L2 error for the survey dataset is the same and reaches its minimum for both $\epsilon$ of 5 and 10, with the value of 0.102.

Interestingly, while we expect errors to decrease as epsilon increases (due to less noise being added), the trend is not strictly monotonic. For example, epsilon = 5 produced the lowest L1 error for the survey dataset and the lowest L2 error for both datasets. This inconsistency could be due to the complexity of AIM's internal optimization process, which prioritizes certain query workloads over others, uses regularization during learning, or inherently favors some marginal combinations in the synthetic generation process. It's worth noting that we focused on 2-way marginals specifically because AIM, in its default configuration, is designed to optimize and generate data using workload queries of size 2.

\section{Discussion}
\label{sec:discussion}


In this study, we generated differentially private synthetic versions of a real-world health dataset using the AIM synthesizer and evaluated them across a range of privacy budgets. We examined how well each synthetic dataset preserved utility through multiple metrics. Our findings show that while higher privacy (lower $\epsilon$) introduces noticeable distortion, a mid-range privacy budget ($\epsilon$ = 5 or 10) achieves a promising balance, offering meaningful utility while still maintaining strong privacy protections.

\subsection{Utility-Privacy Trade-off Analysis}

DP offers a tunable parameter, $\epsilon$, to balance privacy protection and data utility. Lower $\epsilon$ values introduce more noise, enhancing privacy but potentially distorting analytical insights. Conversely, higher values reduce noise, improving utility at the expense of privacy. However, Rosenblatt et al. \cite{rosenblatt2024epistemic} argue that privacy cannot be reduced to a single numeric threshold, such as L1 or L2 errors. Its meaning and significance depend on the context of data use. In this work, we adopt this perspective and empirically investigated how varying $\epsilon$ impacts downstream analyses in the context of health data.

Our findings indicate that an $\epsilon = 5$ strikes a balance: it ensures adequate privacy while preserving essential statistical properties. As visualized in our regression models, UMAP projections, Spearman correlation heatmaps, L1, and L2 error distance measurements, $\epsilon = 5$ retains structural relationships and predictive power without explicitly revealing individual data points. Lower values, such as $\epsilon = 1$, resulted in significant utility degradation, while higher values offered higher utility at the cost of reduced privacy protections. This supports the notion that DP should be evaluated based on its practical performance in context-specific use cases.

\subsection{Effectiveness of AIM}

Our experiments demonstrate the efficacy of AIM in generating private synthetic health data. AIM successfully generates datasets that retain utility across regression tasks and correlation structure, even under moderate privacy budgets. For instance, at $\epsilon=5$, AIM-generated synthetic dataset maintained a close enough $R^2$ score in a stress prediction model and produced correlation heatmap and UMAP structures that closely resembled the original data.

DP, and tools like AIM, offer an auditable framework that organizations can adopt to support their privacy claims with formal guarantees. Incorporating DP into real-world data workflows could transform how privacy policies are written and evaluated, making it possible to verify whether data sharing practices align with stated protections. This shift not only improves transparency but also strengthens public trust in institutions handling sensitive data.

The workload-aware nature of AIM, which focuses on specific marginals relevant to the analysis, contributes to its strong performance. By concentrating the privacy budget on important relationships, it preserves meaningful patterns while minimizing the exposure of individual-level data. This makes AIM particularly suitable for health and behavioral datasets, where maintaining statistical trends is essential for the impact of research, but privacy risks are acute.

While our results highlight $\epsilon=5$ as a good balance for this dataset and use case, we do not recommend a one-size-fits-all value. Rather, we argue for the feasibility and importance of evaluating multiple privacy budgets depending on the analytic goals and sensitivity of the variables involved. Although such evaluations may require computational resources, especially for iterative or workload-aware mechanisms like AIM, they enable a more tailored and rigorous selection of $\epsilon$ values aligned with real-world utility and risk trade-offs.

\subsection{Limitations}

While our study highlights the strengths of AIM, its ability to produce utility-preserving synthetic datasets, maintain statistical relationships under moderate privacy budgets, and support privacy claims with formal guarantees, it is not without limitations. Firstly, even a moderate privacy budget of $\epsilon = 5$ may be deemed excessive in certain regulatory or clinical settings, particularly when conducting a critical life-dependent study whose outcome is intended to predict a dose of critical medication \cite{fredrikson2014privacy}.

Secondly, while AIM preserves some pairwise relationships, maintaining complex multivariate correlations remains challenging, especially for high-dimensional datasets. Additionally, AIM is designed exclusively to preserve numerical attributes and disregards qualitative values.
Lastly, we did not assess re-identification risks beyond basic linkage attacks. Future work could explore more sophisticated adversarial models and examine the impact of DP on mitigating adversarial attacks on the private dataset.

\section{Conclusion}

In this work, we evaluated the performance of the AIM algorithm for generating differentially private synthetic health data using a real-world dataset on college students' mental health. We compared outputs across a range of $\epsilon$ values and found that $\epsilon=5$ offered the best balance of privacy and utility. This aligns with recent scholarship that emphasizes that privacy evaluation must consider the practical context and goals of data use.

Future work will explore alternative DP mechanisms such as Private-PGM variants and GAN-based approaches, as well as utility preservation at lower privacy budgets (e.g., $\epsilon<1$). We also plan to assess the robustness of these mechanisms under more advanced privacy attacks and deploy them in practical research workflows involving underrepresented groups, where privacy risks are especially pronounced.

\bibliographystyle{IEEEtran}
\bibliography{references}

\end{document}